\documentclass[11pt,dvips]{article}
\usepackage{graphicx}
\usepackage{epsfig}

\begin{document}
\title{{\bf {\Large Critical Velocity for He II Superfluidity Motion  \\[-0.15cm] 
in Channels Narrower than Micrometer}}}
\author{J.X. Zheng-Johansson
\\
 {\small {\it   Institute of Fundamental Physics Research,  61193 Nyk\"oping, Sweden}}
\\[0.1cm] 
{\small October, 2004; updated August, 2006; June, 2009}
} 
\date{}

\maketitle

\def\ef{{\rm ef}}
\def\uex{{\rm u}}
\def\ev{\varepsilon}
\def\vdw{{\rm vdw}}
\def\mef{m_{_{\rm {ef}}}}
\def\ov{\over}
\def\lf{\left}
\def\rt{\right}
\def\ph{{\rm ph}}
\def\hquad{\ \ }

\begin{abstract}
We derive an exact equation for the critical velocity for He II superfluidity motion in channels of width $<10^{-5}$ m for which no quantitatively satisfactory prediction exists prior to this work. 
\\
\\
{\scriptsize  *This e-print version (2) contains fuller list of the earlier reports on the new superfluid theory.} 
\end{abstract}

\section*{}

\begin{figure}[t]
\centering
\includegraphics[width=0.75\textwidth]{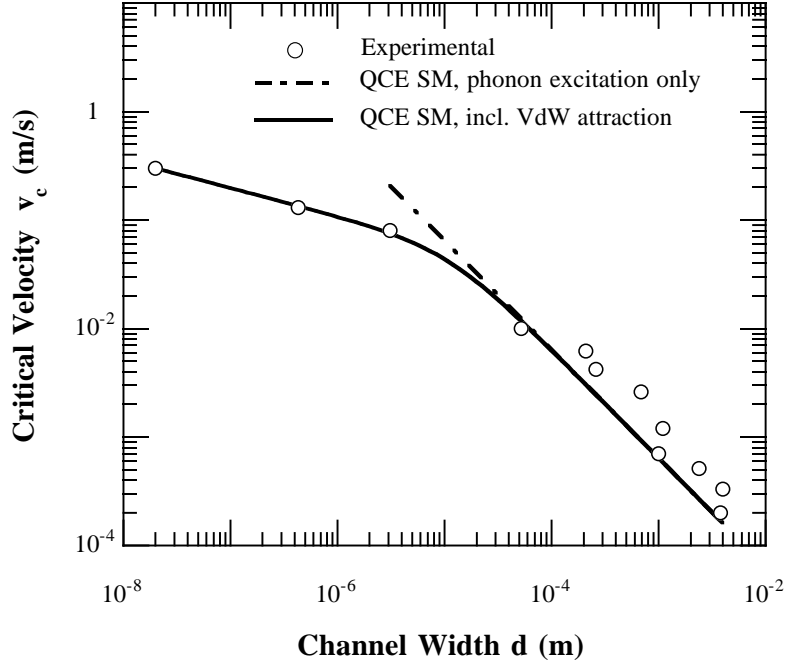}
\caption{Critical velocity $v_c(d)$ versus channel width $d$ for superfluid He II. Circles represent experimental data at about 1.4 K (compiled by J. Wilks, {\it The Properties of Liquid and Solid Helium}, Clarendon Press,  Oxford, 1967, page 391, Table 1.)   
Broken line shows QCE theoretical result  (\ref{eq3-7c}) including phonon excitations only.  
Solid line shows QCE theoretical result (\ref{eq3-7ca}) including a fluid-wall van der Waals attraction in addition to the basic phonon excitations.   The solid line coincides with the broken line in the high $d$ end, implying a negligible 
van der Waals attraction here.
}
\label{fig-vc}
\end{figure}

In channels of width $d>10^{-5}$ m, the critical velocity for He II superfluidity motion can be satisfactorily described  by\cite{Sup2,Sup1} 
$$\displaylines{ \refstepcounter{equation}  \label{eq3-7c}
\hfill  v_c (d) = \sqrt{{48 hc_1 a \over  m  }} \ \frac{1}{d}, 
\hfill (\ref{eq3-7c})
}$$
a prediction based on the QCE (quantum confinement effect) superfluidity mechanism.  
In (\ref{eq3-7c}), $a$ ($\sim 3.6 \times 10^{-10}$ m) is the average interatomic separation distance of He II, $h$ Planck constant,  $m$ ($6.64 \times 10^{-27}$ kg) helium 4 atomic mass, and $c_1$ ($239$ m/s) first sound velocity in He II. 
Substituting with the given values above into   (\ref{eq3-7c}) gives 
$v_c(d)=6.42\times10^{-7}\frac{1}{d}$ m/s, as graphically shown by the broken line in Figure \ref{fig-vc}.
  A similar formula to (\ref{eq3-7c}) was earlier   derived by R. Feynman (1954) on an alternative theoretical argument.

For $d<10^{-5} $m,  $v_c(d)$ of (\ref{eq3-7c}) deviates significantly from the experimental data, as the broken line of Figure \ref{fig-vc} clearly shows.   
Recent inelastic neutrons scattering experiments \cite{Sokol:etal2002,Fak-PRL2000} have shown that, except for some layers by the wall, the bulk excitation spectrum of He II is not altered because of confinement even at a scale of $\sim 70$ \AA. 
So a the phonon excitation picture of He II for deriving (\ref{eq3-7c}) apparently retains valid for the low $d$ region down to plausibly a few $a$. 
On the other hand, we observe that as $d$ reduces, yet with $d>>a$, the dissipation, say of energy $\ev_{\vdw}$ per interfacial atom, 
due to the fluid-wall van der Waals attraction becomes increasingly significant compared to the (reduced) flow energy, $\ev_s$, as a result that the atomic population near the wall becomes increasingly a significant fraction compared to the atomic population in the entire fluid. This is contrasted to in the wider channels where $\ev_{\vdw}<<\ev_s$; $\ev_{\vdw}=0$ is one basic assumption for deriving (\ref{eq3-7c}). 
We below reexpress the critical velocity by including the VdW dissipation. 

Suppose the total VdW force acted by the wall atoms  on the flow atoms as mapped to per interfacial-layer atom is $f$.  
The presence of the VdW force, for its given characteristics, should not alter the underlying QCE mechanism for  the occurrence of a critical velocity\cite{Sup1,Sup2}.  
Consider the flow  has just reached its critical velocity $v_s=v_c$ at time $t=0$, and is then decelerated to $v_s=0$ at $t=\delta t$ due to the total dissipative force on it. In this course, the dissipative work done by the VdW force between the wall on the flow in $\delta t$  is:
$$\displaylines{ \refstepcounter{equation}   \label{eq-eq-vdw2}
\hfill 
\ev_{\vdw} = \int_{ 0}^a f d \ell = \int_{ v_c}^0 m_\vdw v_s  dv_s = -\frac{1}{2} m_\vdw v_c^2 = -\frac{1}{2} (f\delta t) v_c.   \hfill (\ref{eq-eq-vdw2})
}$$
 $$\displaylines{ \refstepcounter{equation}   \label{eq-ft}
{\rm where }\hfill f \delta t = m_\vdw v_c 
                  \hfill (\ref{eq-ft})
}$$  
is the flow momentum consumed in time $\delta t$; $ m_\vdw $ is the effective fluid mass per interfacial-layer atom the VdW force does work on.  
$f$ may be in turn expressed based on two phenomenological considerations: 
 (a)  $f$ is as just said the sum of the VdW attractions of all of the wall atoms with all of the fluid atoms $N_s$, mapped on to per collision atom out of a total $N_A$ collision atoms, thus $f\propto N_s/N_A$. Or equivalently, 
$f\propto N_s m/N_A=M_s/N_A = \mef$, where 
$$\displaylines{\refstepcounter{equation}
\label{eq45ef}
\hfill \mef=(N_s/N_A)m=md/4a \hfill (\ref{eq45ef})
}$$
 (b) From general condensed matter theory we know that, $f \propto 1/r^6$ for  VdW attraction. 
This function is short ranged; so atoms far away from the wall, say beyond a characteristic distance, $d_0$, will not contribute to $f$.  
And for $d<d_0$, the smaller the width $d$, the greater portion of the flow atoms will be subject to the VdW attraction from the wall. Hence,  $f \propto (d_0/d)^\uex$; $\uex>0$. The exponent $\uex$  
reflects the net effect of the influences of (a)--(b).
 $(d_0/d)^\uex$ scales $\mef $  of item (a) as 
$$  \refstepcounter{equation}   \label{eq-f}
f \cdot (\delta t /v_c)= \mef (d_0/d)^\uex.  
\eqno(\ref{eq-f})$$
The factor $ (\delta t /v_c)$ is multiplied to the left-hand side of (\ref{eq-f})  to yield a correct dimension as according to (\ref{eq-ft}).
The equality of the two sides of (\ref{eq-f}) will be ensured by a realistic choice of the values of $d_0$ and $u$ below.
Using (\ref{eq-f}),  (\ref{eq-eq-vdw2}) writes:  
 $$ \ev_{\vdw} =-{1\ov 2} \mef  \lf(\frac{d_0}{d}\rt)^\uex v_c^2.        
                      \eqno(\ref{eq-eq-vdw2})'
$$
Comparing (\ref{eq-eq-vdw2})$'$ and (\ref{eq-eq-vdw2}) we have $m_\vdw=  \mef  \lf(\frac{d_0}{d}\rt)^\uex$. 

If $f=0$, then a given external energy supply would drive the flow into a velocity $v_s'=v_c'$ at time $t=0$, with a reduced critical flow energy ${\ev_s^c}'= {1\ov 2}\mef v_c'^2$. 
But now assuming $f \ne 0$, the same external energy will thus only drive the flow velocity to  $v_s=v_c$ and a reduced critical flow energy 
$$\refstepcounter{equation}\label{eq-Es}
\ev_s^c= {1\ov 2}\mef v_c^2 
                \eqno(\ref{eq-Es})$$ 
The two flow energies are related by: 
$$ {1\ov 2} \mef {v_c'}^2- \ev_{\vdw}  = {1\ov 2} \mef {v_c}^2.
$$

Swapping $\ev_{\vdw} $ to the right side, substituting  (\ref{eq-eq-vdw2})$'$ for it,  and reorganizing, we get: 
$$  
 {\ev_s^c}'= {1\ov 2} \mef {v'_c}^2  = {1\over 2} \mef   \lf(1+ \lf(\frac{d_0}{d}\rt)^\uex \rt) {v_c}^2 =  {1\over 2} \mef' {v_c}^2.   \eqno(\ref{eq-Es})'$$
Where, 
$$
\mef'= \mef  \lf(1+ \lf(\frac{d_0}{d}\rt)^\uex \rt)          \eqno(\ref{eq45ef})' 
$$
represents a new effective mass and has a clear physical meaning. Namely, due to the VdW attraction from the wall, the fluid has a larger effective mass $\mef'$; $\mef'$ increases with a decreasing $d$. 

The threshold condition for the translational to thermal energy conversion for per collision atom is\cite{Sup2,Sup1}:
$\ev_s^{c} (={\ev_s}_{{\rm min}})= \ev_{\ph}^{c}$, where $\ev_{\ph}^{c}=6hc_1/d$. 
Replacing now ${\ev_{s}^{c}}$ of the equation above by the appropriate ${\ev_{s}^{c}}' $, we get the threshold condition for the fluid translational to thermal energy conversion per collision atom in the presence of a fluid-wall VdW attraction:  
$$\refstepcounter{equation} \label{eq3-7b1}
{\ev_s'}^{c}= \ev_{\ph}^{c}.   \eqno(\ref{eq3-7b1})$$

Substituting  (\ref{eq-Es})$'$ for ${\ev_s'}^{c}$ and  $6hc_1/d$ for $\ev_{\ph}^{c}$ as just given
into (\ref{eq3-7b1}), rearranging, we obtain the corresponding critical velocity:
$$ \refstepcounter{equation} \label{eq3-7ca}
v_c = \sqrt{{48 h c_1 a \ov m}} \ {1 \ov d} \frac{1}{\sqrt{1+ \lf(\frac{d_0}{d}\rt)^\uex}}.    \eqno(\ref{eq3-7ca})
$$
\index{Critical velocity, prediction of}
$d_0$ and $\uex$ are adjustable parameters and may be fixed by requiring $v_c(d) $ to agree with the experimental critical velocity data. The results from a least-squares fit (solid line, Figure \ref{fig-vc}) are $d_0= 1.11 \times 10^{-5}$ m and $\uex=1.48$. 
Observe that, the fitted values of $d_0$ and $\uex$ affect {\it only} the bending extent of the curve $v_c(d)$ vs. $d$ in the low $d$ end, which is the characteristic region of item (ii) below. 

$v_c(d)$ divides according to the behavior of $ {1}/{\sqrt{1+ \lf(\frac{d_0}{d}\rt)^\uex}}$ of (\ref{eq3-7ca}) 
  in three characteristic regions: 

{\bf (i).}  $d>>d_0$, thus ${1}/{\sqrt{1+ \lf(\frac{d_0}{d}\rt)^\uex}} \simeq 1$. Here  (\ref{eq3-7ca}) identifies with  (\ref{eq3-7c}); particularly notice that here $v_c(d)$ of  (\ref{eq3-7ca})  is not affected by the values of $d_0$ and $\uex$.  
The above in Figure \ref{fig-vc} corresponds to in the high $d$ end the solid and broken lines coincide. 

{\bf (ii).}  $d<<d_0$, thus $ {1}/{\sqrt{1+ \lf(\frac{d_0}{d}\rt)^\uex}} \simeq (d/d_0)^{\uex/2} <<1$. (\ref{eq3-7ca}) reduces to:  
$$ \displaylines{
\hfill v_c = \sqrt{{48 h c_1 a \ov m}} \ {1 \ov d^{1-\uex/2} d_0^{\uex/2} }  
 = { 2.98 \times 10^{-3}  \over d^{0.26}} \hquad  {\rm m/s}.   \qquad \hfill 
(\ref{eq3-7ca}{\rm a})
}
$$
For the parameterization, values for $d_0$ and $\uex$ are as just given from the fitting, and for the other quantities are as given earlier for parameterizing  (\ref{eq3-7c}). The $v_c(d)$ values of (\ref{eq3-7ca}a) are lowered than given by (\ref{eq3-7c}) for $d<d_0=1.1 \times 10^{-5}$ m, owing to the larger effective fluid mass here produced by the fluid-wall VdW attraction. This in Figure \ref{fig-vc} corresponds to the solid line bending into a flatter one for $d<10^{-5}$ m.

{\bf (iii).} $d\sim d_0$.  Here $v_c(d)$ undergoes a gradual switch from the steeper function of  (\ref{eq3-7c}) to the  flatter  one of (\ref{eq3-7ca}a). As we see in Figure \ref{fig-vc}, $d_0=1.1 \times 10^{-5}$ m represents just the channel width where the steeper broken line starts to give way to the flatter sector of the solid line.

The core part of the helium research, which the author  carried out  as a visiting scientist at the HH Wills Lab, Bristol University, 1998-1999, 
 was supported by the Swedish Natural Science Research Council (NFR) and partly by the Wallenbergs Stiftelse (WS); the research was  indebted to the encouragement from Dr J Wilson, Prof B Johansson, Prof M Springford, Dr P Meeson, Prof K Skold,  and a number of international specialists in liquid helium and superconductivity with whom the author communicated about the work, and to the continued moral support of Senior Scientist P-I Johansson.

\end{document}